\documentclass[aip,reprint,jcp,twocolumn]{revtex4-1}
\usepackage{mhchem}
\usepackage[paperheight=11in,paperwidth=8.25in,margin=0.67in]{geometry}
\usepackage{amsmath}
\usepackage{amssymb}
\usepackage{graphicx,color}
\usepackage{url}
\usepackage{bbm}
\usepackage{algpseudocode}
\usepackage{algorithm}

\usepackage{amsthm}

\newtheorem{thm}{Theorem}

\newcommand{\hH}{\hat{H}}
\newcommand{\hV}{\hat{V}}

\newcommand{\hN}{\hat{N}}

\newcommand{\perm}{\mathfrak{g}}
\newcommand{\sgn}{\mathrm{sgn}}
\newcommand{\rMP}{\mathrm{MP}}
\newcommand{\ii}{\mathbbm{i}}

\begin{document}
\title{
Stochastic many-body perturbation theory for electron correlation energies
}
\author{Zhendong Li}\email{zhendongli2008@gmail.com}
\affiliation{Key Laboratory of Theoretical and Computational Photochemistry, Ministry of Education, College of Chemistry, Beijing Normal University, Beijing 100875, China}

\begin{abstract}
Treating electron correlation more accurately and efficiently is at the heart of the
development of electronic structure methods. In the present work, we explore the use of stochastic approaches to evaluate high-order electron correlation energies, whose conventional computational scaling is unpleasantly steep, being $O(N^{n+3})$ with respect to the system size $N$ and the perturbation order $n$ for the M{\o}ller-Plesset (MP) series.
To this end, starting from Goldstone's time-dependent formulation of \emph{ab initio} many-body perturbation theory (MBPT), we present a reformulation of MBPT,
which naturally leads to a Monte Carlo scheme with $O(nN^2+n^2N+f(n))$ scaling at each step, where $f(n)$ is a function of $n$ depending on the specific numerical scheme.
Proof-of-concept calculations demonstrate that
the proposed quantum Monte Carlo algorithm
successfully extends the previous Monte Carlo approaches for MP2 and MP3 to higher orders
by overcoming the factorial scaling problem. For the first time the Goldstone's time-dependent formulation is made
useful numerically for electron correlation energies, not only being purely as a theoretical tool.
\end{abstract}
\maketitle

\section{Introduction}
The development of accurate and efficient methods for electron correlations
is an enduring frontier in quantum chemistry. For weakly correlated systems, the many-body perturbation theory (MBPT) and the coupled-cluster (CC) theory\cite{shavitt2009many}
have been well-established as the standard tools due to their high accuracy and size extensity.
The later property is essential for large systems.
However, their unfavorable scalings make the application to situations that require
high accuracy still being a significant challenge. Important situations include
the studies of polymorphism of pharmaceutical solids\cite{beran2016modeling}
and relative stabilities of different structures of water clusters\cite{ludwig2001water},
which typically require an accuracy of 0.1kcal/mol.
Although local correlation methods\cite{saebo1993local,schutz1999low,riplinger2013efficient} with reduced scalings have been significantly advanced in recent years, the lack of a method to
benchmark their accuracies for large systems is also a problem to be solved.
Due to its low scaling $O(N^3)$ with $N$ being the system size,
the diffusion quantum Monte Carlo (DMC)\cite{foulkes2001quantum,dubecky2016noncovalent} is gaining
increasing popularity for large systems in recent years.
However, the errors introduced by the fixed node approximation also require careful calibrations.
Therefore, there is clearly still a need for developing methods with guaranteed accuracy for large systems even in the weakly correlated regime.

Motivated by various recent developments of quantum Monte Carlo (QMC) algorithms
in \emph{ab initio} quantum chemistry and condense matter physics
\cite{thom2007stochastic,booth2009fermion,cleland2010communications,
thom2010stochastic,scott2019diagrammatic,
willow2012stochastic,willow2014stochastic,willow2013stochastic,
neuhauser2012expeditious,neuhauser2014breaking,cytter2014metropolis,neuhauser2017stochastic,dou2019stochastic,
sharma2017semistochastic,garniron2017hybrid,guo2018communication,motta2018ab,
rubtsov2004continuous,rubtsov2005continuous,gull2011continuous,rossi2017determinant,van2018diagrammatic},
in the present work, we explore the possibility of using stochastic approaches to evaluate high-order electron correlation energies. Specifically, we will present a reformulation of standard MBPT into a general mathematical form, which is more suitable for Monte Carlo evaluations.
In sharp contrast to the steep scaling $O(N^{n+3})$ of the conventional algorithm\cite{cremer2011moller} for the $n$-th order in the M{\o}ller-Plesset\cite{moller1934note} (MP) series, the resulting QMC algorithm for evaluating the correlation energies scales as $O(N^2)$ with respect to the system size.
It deserves to be emphasized that
although it is well-known that MBPT is less robust than CC and may fail to converge even for simple weakly correlated systems\cite{olsen1996surprising}, having the ability to compute high order MP$n$ energies for large systems is still highly valuable. Because mathematically the convergence issue can be overcome when more information
about the behaviors of the MP series is available. This has been demonstrated for small molecules via resummation techniques\cite{goodson2012resummation}.
In view of its low scaling, the proposed QMC algorithm may potentially open up the possibility
to compute large systems (or small systems but with very large basis sets)
with high accuracy without resorting to any local approximation.

\section{Recapitulation of MBPT}
To begin with, we briefly recapitulate the standard MBPT in Goldstone's
time-dependent formulation\cite{goldstone1957derivation}, which is the
starting point of our reformulation of MP$n$ correlation energies.
For simplicity, we will focus on the MP partition, viz.,
$\hat{H}=\hat{H}_0+\hat{V}$ with $\hat{H}_0=\sum_p\varepsilon_p a_p^\dagger a_p$,
where the zeroth order state is the \emph{canonical} Hartree-Fock (HF) reference $|\Phi_0\rangle=|\Phi_{\mathrm{HF}}\rangle$
and $\hat{V}=\hat{V}_{ee}-\hat{V}_{\mathrm{HF}}$.
The Goldstone's linked-cluster theorem\cite{goldstone1957derivation} states that
the $(n+1)$-th order correlation energy is given by
\begin{widetext}
\begin{eqnarray}
E_{n+1} &=&
\lim_{\epsilon\rightarrow 0}
\frac{(-\ii)^n}{n!}\int_{-\infty}^{0} dt_1\cdots dt_n e^{-\epsilon(|t_1|+\cdots+|t_n|)}
\langle\Phi_0|\hV_I(0)T[\hV_I(t_1)\cdots \hV_I(t_n)]|\Phi_0\rangle_{c},\label{ZT_LCT}
\end{eqnarray}
\end{widetext}
where $\hV_I(t)=e^{\ii\hH_0 t}\hV e^{-\ii\hH_0 t}$ is the corresponding perturbation operator
in the interaction picture and the subscript 'c' indicates the connected parts.
For the Coulomb interaction, $\hat{V}_{ee}$ can be written as
\begin{eqnarray}
\hat{V}_{ee}=\frac{1}{2}\int dx_1dx_1' v(x_1,x_1')\hat{\psi}^\dagger(x_1)\hat{\psi}^\dagger(x_1')\hat{\psi}(x_1')\hat{\psi}(x_1),\label{Coulomb}
\end{eqnarray}
where $x_1\triangleq(\sigma_1,\vec{r}_1)$ and $x_1'\triangleq(\sigma_1',\vec{r}_1')$ are composite indices
for spin and spatial variables, $v(x_1,x_1')=\delta_{\sigma_1\sigma_1'}/|\vec{r}_1-\vec{r}_1'|$,
and the integration over $x_1$ implies both a summations over spin $\sigma_1\in\{\alpha,\beta\}$
and an integration over spatial coordinates $\vec{r}_1\in\mathbb{R}^3$.
For brevity, we also introduce a compact notation $\hat{\psi}^{(\dagger)}(1)\triangleq\hat{\psi}^{(\dagger)}(x_1,t_1)$
to represent operators in the interaction/Heissenberg picture by defining $1\triangleq(t_1,\sigma_1,\vec{r}_1)$.

Since $\hat{H}_0$ is quadratic, the (physical) vacuum expectation value in Eq. \eqref{ZT_LCT} can be evaluated using the Wick's theorem\cite{wick1950evaluation}, which implies a factorization of high-order noninteracting Green's functions into a sum over products of one-body Green's function (propagators)\cite{stefanucci2013nonequilibrium},
\begin{eqnarray}
&&G^0(1,\cdots,n;1',\cdots,n')\nonumber\\
&\triangleq&(-\ii)^n\langle\Phi_0|T[\hat{\psi}(1)\cdots
\hat{\psi}(n)\hat{\psi}^\dagger(n')\cdots\hat{\psi}^\dagger(1')]|\Phi_0\rangle\nonumber\\
&=&\det(\mathbf{G}^0_{n}),\label{Wick}
\end{eqnarray}
where $\mathbf{G}^0_{n}$ is an $n$-by-$n$ matrix with entries $(\mathbf{G}^0_{n})_{kl}\triangleq G^0(k,l)=
-\ii\langle\Phi_0|T[\hat{\psi}(k)\hat{\psi}^\dagger(l)]|\Phi_0\rangle$.
Applying the Wick's theorem \eqref{Wick} in Eq. \eqref{ZT_LCT} and
expanding the determinant, each product can be represented by a Feynman (Goldstone) diagram either connected or disconnected, and the linked cluster theorem\cite{goldstone1957derivation} states that
only the connected parts contribute to $E_{n+1}$. This is the standard way to derive Feynman diagrams
in MBPT. However, for our purpose, we will keep Eq. \eqref{Wick} in
its unexpanded form. The central result of this work is to
show that for the ground state (a more precise condition
will be given in Sec. \ref{sec:imagt}), Eq. \eqref{ZT_LCT} can be recast into the following
general mathematical form, which is more suitable for stochastic evaluations than the form based on
Feynman (Goldstone) diagrams,
\begin{eqnarray}
E_{n+1}=\frac{(-1)^n}{2^{n+1}n!}
\int d\nu_0'd\nu_1 \cdots d\nu_n\;
w_{n+1}\kappa_{n+1}.\label{FinalEq}
\end{eqnarray}
The meaning of notations are explained as follows:
$w_{n+1}\triangleq v_0v_1\cdots v_n$ with the interaction $v_n\triangleq v(x_n,x_n')$,
$\nu_n\triangleq(\tau_n,\sigma_n,\vec{r}_n,\sigma_n',\vec{r}_n')$
is a collection of all coordinates for the pair of electrons
originated from the same $\hat{V}_I(t_n)$, $\tau_n$
is used to differentiate the imaginary time variable from the corresponding real time $t_n$,
see Sec. \ref{sec:imagt}.
The prime in $\nu_0'$ \eqref{FinalEq}
indicates that the integrations over $\nu_0$ exclude the time integration,
since $\tau_0=t_0=0$ from Eq. \eqref{ZT_LCT}. The most important part
$\kappa_{n+1}$ in Eq. \eqref{FinalEq} is a function of all variables
$\kappa_{n+1}(\nu_0,\nu_1,\cdots,\nu_{n+1})$ and will be discussed in Sec. \ref{sec:kappa}.

\section{Imaginary-time formula for zero-temperature correlation energies}\label{sec:imagt}
To recast Eq. \eqref{ZT_LCT} into a form shown in Eq. \eqref{FinalEq}, we proceed in
two steps. First, we show that under certain conditions,
Eq. \eqref{ZT_LCT} can be rewritten in a form similar
to the grand potential $\Omega$ in the grand canonical ensemble
in finite-temperature MBPT\cite{bloch1958developpement,fetter1971quantum} (however,
the subtle difference is discussed in Appendix). The precise statement
is provided by the following theorem.
\begin{thm}[imaginary-time formula]\label{thm:itf}
Assuming the HF reference $|\Phi_0\rangle$ is the ground state of $\hH_0$,
$E_{n+1}$ in Eq. \eqref{ZT_LCT} can be alternatively expressed by
an "imaginary-time" analog, viz.,
\begin{widetext}
\begin{eqnarray}
E_{n+1} &=&
\frac{(-1)^n}{n!}\int_{-\infty}^{0} d\tau_1\cdots d\tau_n
\langle\Phi_0|\hV_I(0)T[\hV_I(\tau_1)\cdots \hV_I(\tau_n)]|\Phi_0\rangle_{c},\label{eFTMBPT}
\end{eqnarray}
\end{widetext}
where $\hV_I(\tau)=e^{\tau (\hH-\mu \hN)}\hV e^{-\tau(\hH-\mu\hN)}$ and the
analog of "chemical potential" $\mu$ is an arbitrary constant here.
\end{thm}
The form \eqref{eFTMBPT} is more computational appealing, since it
only involves real quantities, whereas Eq. \eqref{ZT_LCT}
involves oscillating complex quantities arising from $e^{\ii\hat{H}_0t}$,
which will make the later stochastic evaluation more challenging in general.
Besides, the unpleasant adiabatic factor $e^{-\epsilon|t|}$ is also removed
from Eq. \eqref{ZT_LCT}.

There are several ways to derive Eq. \eqref{eFTMBPT}, but the most obvious way
is to consider the well-known time-independent expression\cite{goldstone1957derivation} derived from
Eq. \eqref{ZT_LCT},
\begin{eqnarray}
E_{n+1}=
(-1)^n\sum_{\{I_i\}}\left(V_{0I_1}\frac{1}{\omega_{I_1}}V_{I_1I_2}\frac{1}{\omega_{I_2}}\cdots \frac{1}{\omega_{I_n}}V_{I_n0}\right)_{c},\label{ZTMBPTtimeIndependent}
\end{eqnarray}
where $V_{I_1I_2}=\langle\Phi_{I_1}|\hV|\Phi_{I_2}\rangle$
with $|\Phi_{I_i}\rangle$ being intermediate states, and
$1/\omega_{I}\triangleq 1/(E_I-E_0)$ represents the energy denominator. By performing
a Laplace transform $1/\omega_{I}=\int_{-\infty}^0 d\tau
e^{\omega_{I}\tau}$ and reversing the derivation
from time-dependent PT \eqref{ZT_LCT} to time-independent PT \eqref{ZTMBPTtimeIndependent},
we can derive Eq. \eqref{eFTMBPT} from Eq. \eqref{ZTMBPTtimeIndependent}.
However, two differences need to be noted.

First, the condition for the Laplace transform requires $\omega_I>0$
for all intermediate states, and hence Eq. \eqref{eFTMBPT} is valid only for the case that $|\Phi_0\rangle$
is the ground state of $\hH_0$, whereas Eq. \eqref{ZT_LCT} based on the
Gell-Mann-Low theorem\cite{gell1951bound} in principle
also works for excited states.
Since we are mainly focused on the ground
state problem in this work, this condition is
usually satisfied.

Second, in Eq. \eqref{eFTMBPT}
we have introduced a parameter $\mu$, which formally corresponds to
the chemical potential in finite temperature MBPT. But here it
is completely arbitrary, as can be seen in Eq. \eqref{ZTMBPTtimeIndependent},
because as long as $\hV$ does not change the particle number,
$\mu$ will be exactly cancelled in taking energy differences in the denominator $1/\omega_{I}$.
However, for numerical convenience, we can choose it to be a value
within the gap between the HOMO (highest occupied molecular orbital) and the LUMO
(lowest unoccupied molecular orbital), e.g., $\mu=\frac{\varepsilon_{\mathrm{HOMO}}+
\varepsilon_{\mathrm{LUMO}}}{2}$ used in the present work, such that given
$|\Phi_0\rangle$, we have $\varepsilon_i'
\triangleq\varepsilon_i-\mu<0$ for occupied orbitals and $\varepsilon_a'\triangleq
\varepsilon_a-\mu>0$ for virtual orbitals. This will make
the exponential factor in the following imaginary time Green's function $G^0(k,l)$,
appearing in the counterpart of Eq. \eqref{Wick}, always smaller than one,
\begin{eqnarray}
G^0(k,l)&=&\theta(\tau_{kl})G^0_>(k,l)+\theta(-\tau_{kl})G^0_<(k,l),\nonumber\\
G^0_>(k,l)&=&-\delta_{\sigma_k\sigma_l}\sum_{a}e^{-\varepsilon_a'\tau_{kl}}
\psi_{a\sigma_k}(\vec{r}_k)\psi_{a\sigma_k}^*(\vec{r}_l)\nonumber\\
G^0_<(k,l)&=&\delta_{\sigma_k\sigma_l}\sum_{i}e^{-\varepsilon_i'\tau_{kl}}
\psi_{i\sigma_k}(\vec{r}_k)\psi_{i\sigma_k}^*(\vec{r}_l),\label{GFtImagTime}
\end{eqnarray}
where $\tau_{kl}\triangleq\tau_k-\tau_l$. The same trick was previously
introduced in the Laplace-transformed MP2\cite{haser1992laplace}.

Alternatively, the above results can be derived from Eq. \eqref{ZT_LCT} by analyzing each \emph{connected} Goldstone diagram and performing an analytic continuation
of the real time integration to the imaginary time.
We will not go into the details, but just mention that
the same condition $\omega_I>0$ in this case will come from the requirement
to guarantee that the contour integral over the arc goes to zero.

\section{Summation of diagrams by
moment-cumulant relations}\label{sec:kappa}
The next step is to express Eq. \eqref{eFTMBPT} into Eq. \eqref{FinalEq}.
Applying the imaginary-time analogy of the Wick's theorem \eqref{Wick} in
Eq. \eqref{eFTMBPT}, the expectation value $\langle\Phi_0|\hV_I(0)T[\hV_I(\tau_1)\cdots \hV_I(\tau_n)]|\Phi_0\rangle_{c}$ for $E_{n+1}$ will become $\frac{1}{2^{n+1}}w_{n+1}\det_c(\mathbf{G}^0_{2n+2})$, where again
the subscript 'c' is used to denote the connected contributions.
For small $n$, $\det_c(\mathbf{G}^0_{2n+2})$ can be explicitly expanded,
which corresponds to the use of Goldstone diagrams as employed in MC-MP2 (Monte Carlo MP2) and MC-MP3 by Hirata et al.\cite{willow2012stochastic,willow2014stochastic}.
However, this approach quickly becomes inefficient as $n$ increases, since the number of diagrams increases \emph{factorially}. While there are only 2 diagrams for MP2 and 12 diagrams for MP3,
MP4 and MP5 have 300 and 13680 Goldstone diagrams\cite{rossky1976enumeration,wilson1985diagrammatic,kucharski1986fifth}, respectively.
Recently, in the context of diagrammatic Monte Carlo,
which samples all Feynman diagrams stochastically,
a trick to sum all connected diagrams at order $n$ was proposed
by Rossi\cite{rossi2017determinant}, by recursively subtracting disconnected
contributions from determinants containing all diagrams.
It has an exponential scaling $O(2^nn^3+3^n)$, but is less than factorial
and has allowed to sum diagrams at order as high as 10 for the Hubbard model\cite{rossi2017determinant}.
The same recursive formula (vide post) can be applied to compute $\det_c(\mathbf{G}^0_{2n+2})$
from $\det(\mathbf{G}^0_{2n+2})$ for correlation energies in Eq. \eqref{eFTMBPT}.
In the following context, we provide a different derivation, which
is more explicit and unveils the underlying fundamental moment-cumulant
relation. More importantly, in this way we are able to write down an explicit expression
for $\det_c(\mathbf{G}^0_{2n+2})$ in terms of principal minors of
$\det(\mathbf{G}^0_{2n+2})$.

\begin{figure}\centering
\begin{tabular}{cc}
\resizebox{!}{0.1\textheight}{\includegraphics{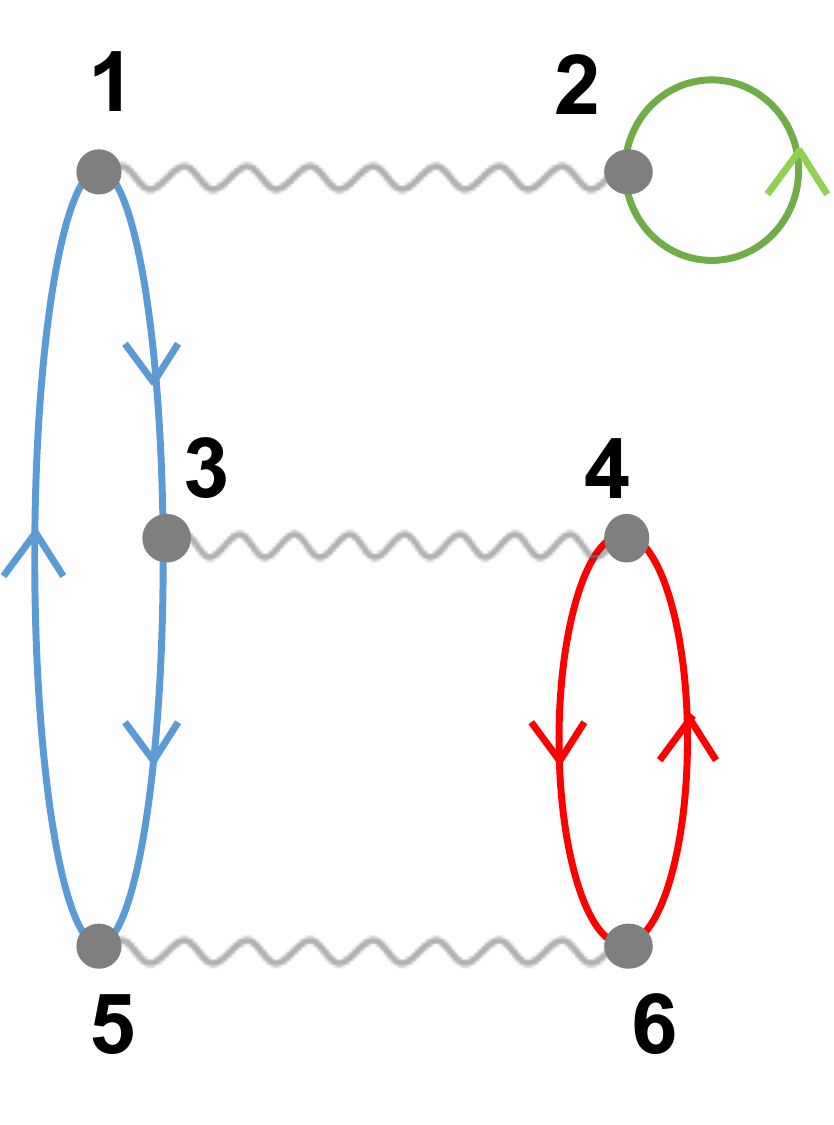}} &
\resizebox{!}{0.1\textheight}{\includegraphics{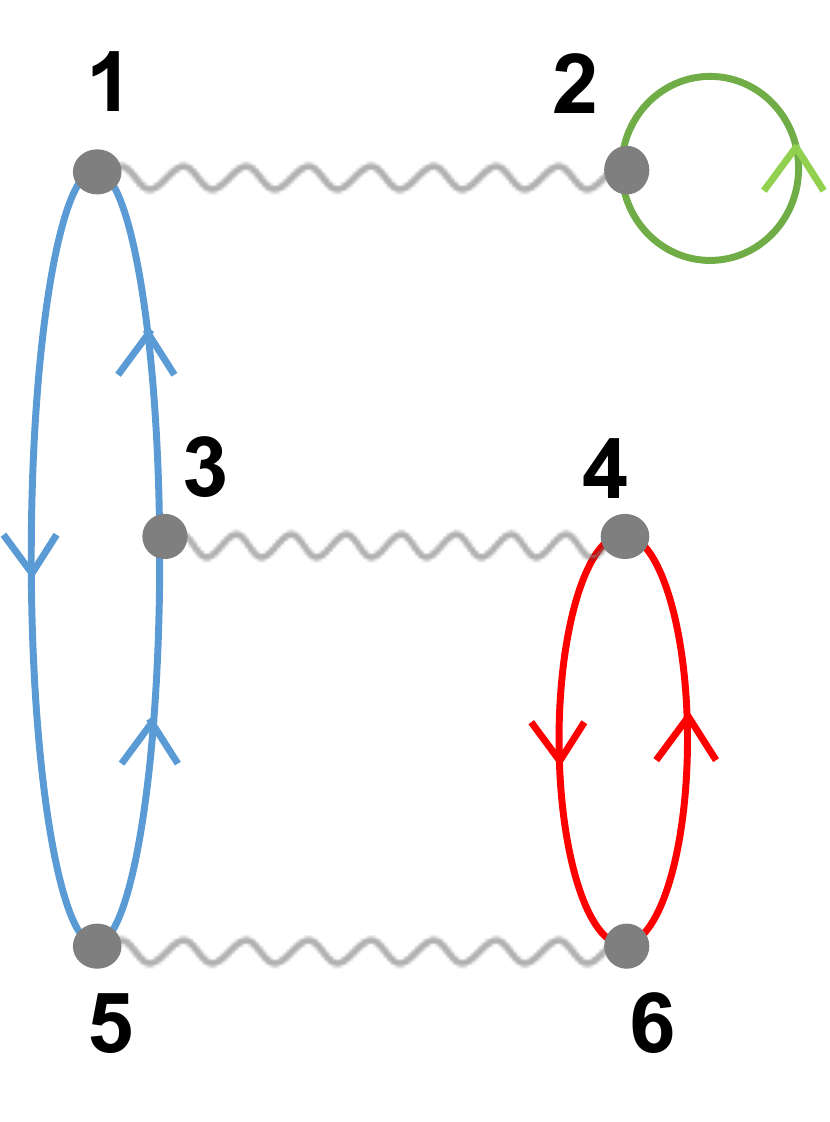}} \\
(a) $\perm_a=(135)(2)(46)$ &
(b) $\perm_b=(153)(2)(46)$ \\
\resizebox{!}{0.1\textheight}{\includegraphics{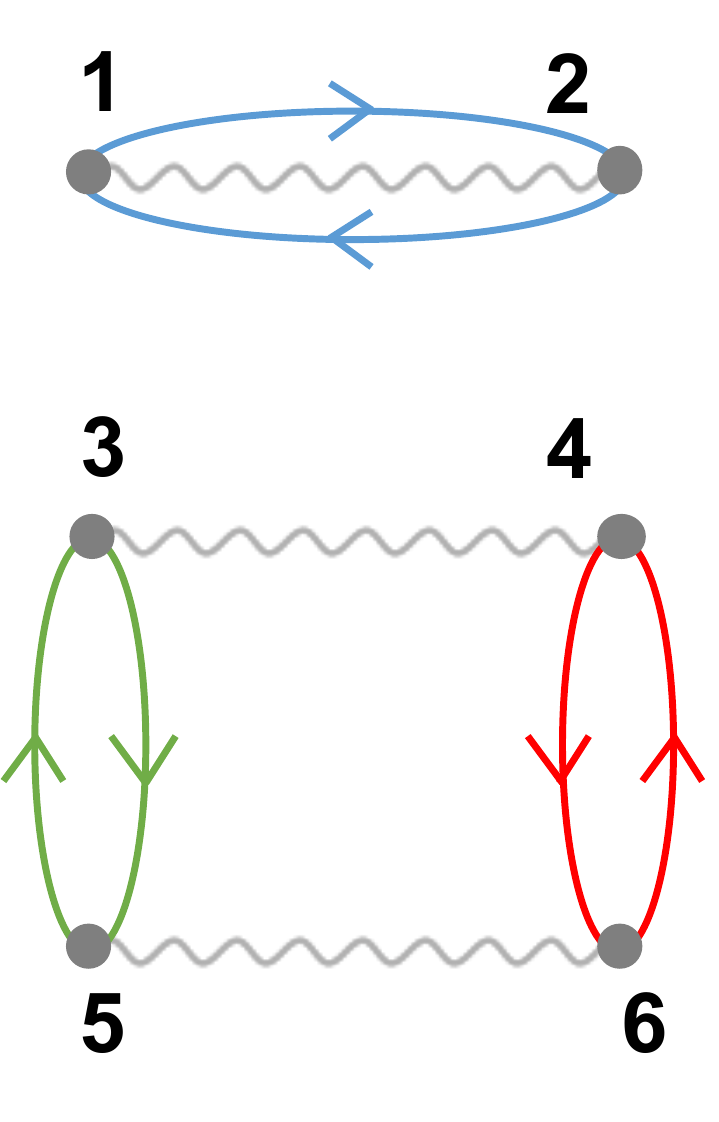}} &
\resizebox{!}{0.1\textheight}{\includegraphics{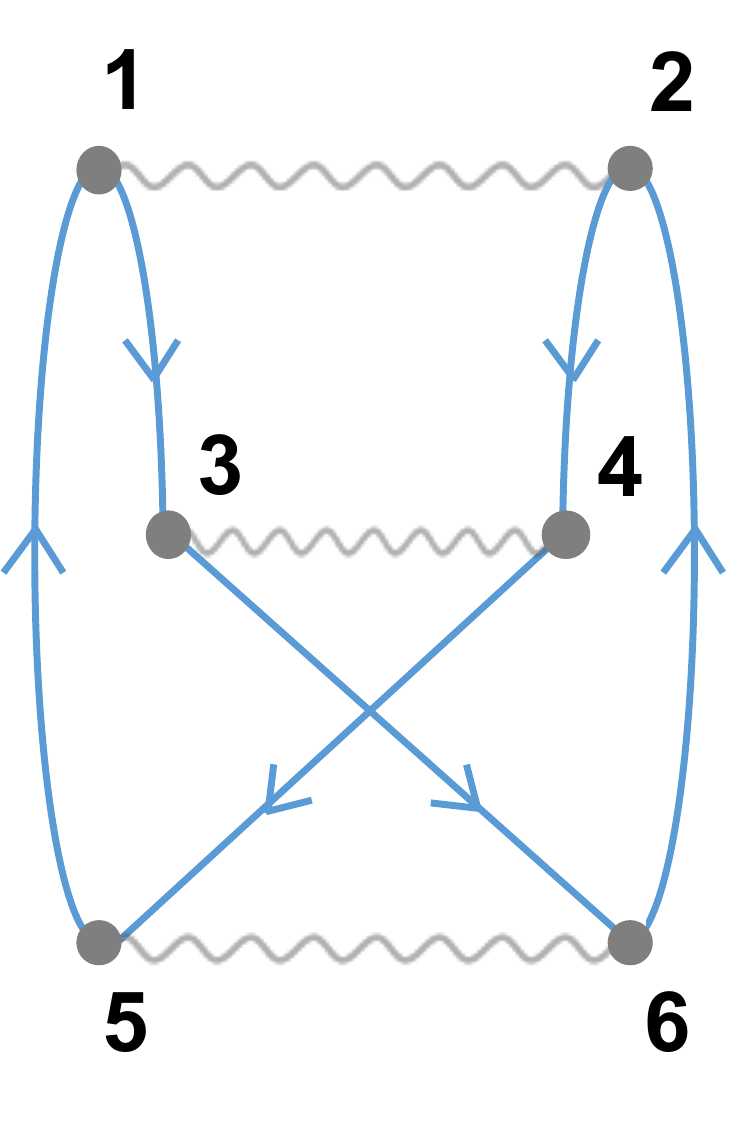}} \\
(c) $\perm_c=(12)(35)(46)$ &
(d) $\perm_d=(136245)$ \\
\end{tabular}
\caption{Examples for cycle decomposition of permutations $\perm\in S_{6}$
and the loop structures of diagrams in MBPT. The loops formed by Green's functions
highlighted by different colors are the basic building blocks, and they
are glued together by interaction lines (wiggles) to form connected
or disconnected diagrams.}\label{fig:cycleDecomp}
\end{figure}

\subsubsection{Loop expansion of determinants}
To extract the connected part $\det_c(\mathbf{G}^0_{2n+2})$ from
$\det(\mathbf{G}^0_{2n+2})$, let us first consider the explicit formula for a determinant $\det(\mathbf{A}_{n})$ of an $n$-by-$n$ matrix $\mathbf{A}_{n}$,
\begin{eqnarray}
\det(\mathbf{A}_{n})=\sum_{\perm\in S_{n}} \sgn(\perm)A_{1,\perm(1)}
\cdots A_{n,\perm(n)},\label{detDecomp}
\end{eqnarray}
where the summation is over all permutations $\mathfrak{g}$ ($\in S_n$) and $\sgn(\perm)$ is
the signature of the permutation $\perm$. For each permutation, we apply the
cycle decomposition, e.g.,
\begin{eqnarray}
\perm_a=\left(\begin{array}{cccccc}
1 & 2 & 3 & 4 & 5 & 6 \\
3 & 2 & 5 & 6 & 3 & 4 \\
\end{array}\right) =  (135)(2)(46).\label{gexample}
\end{eqnarray}
Some typical examples are shown in Fig. \ref{fig:cycleDecomp} for $n=6$.
Since the signature $\sgn(\perm)$ can also be decomposed, e.g.,
$\sgn(\perm_a)=\sgn((135))\sgn((2))\sgn((46))=(-1)^{|(135)|-1}(-1)^{|(2)|-1}(-1)^{|(46)|-1}$,
where $|(135)|=3$ represents the length of the cycle (135), diagrammatically, we are able to rewrite each term in Eq. \eqref{detDecomp} as a product of \emph{loops}, see Fig. \ref{fig:cycleDecomp}.
For example, the term for $\perm_a$ can be written compactly as $l(135)l(2)l(46)$,
where the loop product $l(i_1i_2\cdots i_k)\triangleq (-1)^{k-1}A_{i_1,i_2}\cdots A_{i_{k-1},i_{k}}A_{i_{k},i_1}$
for each cycle $(i_1i_2\cdots i_k)$. Then, the sum over $n!$ permutations in
Eq. \eqref{detDecomp} can be greatly simplified, by realizing that the sum over all the
terms corresponding to the same set partition can be simplified into a single product, e.g.,
$(l(135)+l(153))l(2)l(46)\triangleq\kappa(\{1,3,5\})\kappa(\{2\})\kappa(\{4,6\})$ for permutations $\perm_a$ and $\perm_b$
(see Fig. \ref{fig:cycleDecomp}) generated from the set partition
$\pi=\{\{1,3,5\},\{2\},\{4,6\}\}$.
Thus, through this construction, in general, we can express a determinant
as a sum over partitions, where each partition
contributes to a single product.
\begin{thm}[loop expansion of determinants]
Given an elementary index set $\mathcal{I}=\{1,\cdots,n\}$, Eq. \eqref{detDecomp} can be
re-expressed as
\begin{eqnarray}
\det(\mathbf{A}_{n})&=&\sum_{\pi} \prod_{I_k\in\pi}\kappa(I_k),\nonumber\\
\kappa(I_k) &=& \sum_{i=1}^{(|I_k|-1)!}l_i(I_k).\label{detLoop}
\end{eqnarray}
where $\pi$ represents a \emph{set partition} of
with length $|\pi|$, viz., $\pi = \{I_1,I_2,\cdots, I_{|\pi|}\}$ with the $k$-th
block $I_k=\{i_1,i_2,\cdots,i_{|I_k|}\}$, and $\kappa(I_k)$
is a sum over contributions from the $(|I_k|-1)!$ possible cycles
generated from the index set $I_k$, with $l_i(I_k)$ being the loop
contribution from one of the cycles.
\end{thm}

In a Feynman (Goldstone) diagrammatic language,
Eq. \eqref{detLoop} is nothing but a mathematical description of the fact
that loops formed by Green's functions are building block for all diagrams,
as highlighted by different colors in Fig. \ref{fig:cycleDecomp} for selected third-order diagrams.
Furthermore, the quantity $\kappa(I_k)$ sums over all possible loops generated from
the points in $I_k$. Therefore, if one-body operators are considered as the only perturbations,
then $\kappa(\mathcal{I})$ is simply the sum over all connected diagrams,
because in this case the definition of connectivity with respect to perturbations coincides
with the graphical definition. For two-body operators \eqref{Coulomb},
the connectivity in $\det_c(\mathbf{G}^0_{2n+2})$, defined
with respect to the interaction lines (wiggles in Fig. \ref{fig:cycleDecomp}),
is different from that purely for Green's function lines.
Fortunately, this complication can be treated by a generalized definition of $\kappa$
in the next section.

\subsubsection{Explicit expressions for $\kappa$ in Eq. \eqref{FinalEq}}
For two-body perturbations, where different loops
can be glued together by interaction lines to form a single connected diagrams,
we can simple redefine the index set, viz., $\mathcal{I}=\{\nu_0,\cdots,\nu_n\}$
for $\det(\mathbf{G}^0_{2n+2})$, with each element corresponding
to one interaction pair (i.e., two points in a graph).
Then, it can be verified that exactly the same relation \eqref{detLoop} also holds for this
new index set $\mathcal{I}$.
We illustrate this for the case $\mathcal{I}=\{\nu_0,\nu_1\}$ with
$\nu_0=\{1,2\}$ and $\nu_1=\{3,4\}$.
By defining generalized quantities
$\kappa(\{\nu_0\})\equiv\kappa(\{\{1,2\}\})$ and
$\kappa(\{\nu_0,\nu_1\})\equiv\kappa(\{\{1,2\},\{3,4\}\})$ as
{\small\begin{eqnarray}
\kappa(\{\{1,2\}\}) &\triangleq& \kappa(\{1,2\}) + \kappa(\{1\})\kappa(\{2\}),\\
\kappa(\{\{1,2\},\{3,4\}\}) &\triangleq& \kappa(\{1,2,3,4\}) \nonumber\\
	  &+& \kappa(\{1\})\kappa(\{2,3,4\}) + \kappa(\{2\})\kappa(\{1,3,4\}) \nonumber\\
	  &+& \kappa(\{3\})\kappa(\{1,3,4\}) + \kappa(\{4\})\kappa(\{1,2,4\}) \nonumber\\
	  &+& \kappa(\{1,3\})\kappa(\{2,4\}) + \kappa(\{1,4\})\kappa(\{2,3\}) \nonumber\\
	  &+& \kappa(\{1,3\})\kappa(\{2\})\kappa(\{4\}) \nonumber\\
      &+& \kappa(\{1,4\})\kappa(\{2\})\kappa(\{3\}) \nonumber\\
	  &+& \kappa(\{2,4\})\kappa(\{1\})\kappa(\{3\}) \nonumber\\
      &+& \kappa(\{2,3\})\kappa(\{1\})\kappa(\{4\}),
\end{eqnarray}}where $\kappa$ on the right hand sides
are defined in Eq. \eqref{detLoop} for the elementary
index set $\{1,2,3,4\}$, the determinant $\det(\mathbf{A}_4)$ can be
rewritten as $\det(\mathbf{A}_4)=
\kappa(\{\nu_0,\nu_1\})+\kappa(\{\nu_0\})
\kappa(\{\nu_1\})$. It is of the same form as Eq. \eqref{detLoop}, but for partitions
of $\mathcal{I}$, and $\kappa(\{\nu_0,\nu_1\})$ now represents
the target connected quantity $\det_c(\mathbf{G}^0_{2n+2})$.
The importance of the form \eqref{detLoop} lies in that it is the same as
the moment-cumulant relation in the multivariate case, such that the inversion of this relation is known.
Summarizing these results, we have the following explicit
expression for $\det_c(\mathbf{G}^0_{2n+2})$.
\begin{thm}[moment-cumulant relation]\label{thm:mcr}
Let $\mathcal{I}=\{\nu_0,\cdots,\nu_n\}$,
the moments defined as $\mu(\mathcal{I})\triangleq\det(\mathbf{G}^0_{2n+2})$,
and other $\mu(I_k)$ being the principal minors of $\det(\mathbf{G}^0_{2n+2})$
with both columns and rows constructed from $I_k$, the cumulant
$\kappa(\mathcal{I})\triangleq\det_c(\mathbf{G}^0_{2n+2})$ is given explicitly as
\begin{eqnarray}
\kappa(\mathcal{I})=\sum_{\pi}(|\pi|-1)!
(-1)^{|\pi|-1}\prod_{I_k\in\pi}\mu(I_k),\label{kappamu}
\end{eqnarray}
which can be viewed as an inversion of the relation \eqref{detLoop}
in the particular setting.
\end{thm}
It deserves to point out that the coefficient $(|\pi|-1)!$ is nontrivial in the sense that it is a reflection
of the nontrivial symmetric factors in diagrams for energy/free-energy,
which are more complicated than those in diagrams for Green's functions (which would be simply one).
The connection to the moment-cumulant relation is physically quite appealing, since
it is a reflection of the linked cluster theorem\cite{goldstone1957derivation},
and ensures that the correlation energy at each order is size-extensive.

\subsubsection{Lowest order $\kappa^{\rMP n}$}
The number of partitions in the sum \eqref{kappamu} is given by the Bell number $B_n$, which are $B_2=2$, $B_3=5$, $B_4=15$, and $B_5=52$ for the lowest few orders, and it growth, bound by
$(0.792n/\ln(n+1))^n$\cite{berend2010improved}, is much slower than factorial.
With the HF reference as in our case, Eq. \eqref{kappamu} can be further simplified, since
the effect of the term $-\hV_{\mathrm{HF}}$ in $\hV$ is
equivalent to set the diagonal 2-by-2 blocks of $\mathbf{G}_{2n+2}^0$
be zero, such that $\kappa(\{\nu_i\})=\mu(\{\nu_i\})=0$.
Consequently, the lowest few orders can be expressed compactly as
\begin{eqnarray}
\kappa^{\rMP 2}&=&\mu(\{\nu_0,\nu_1\})\triangleq \mu_{01},\\
\kappa^{\rMP 3}&=&\mu(\{\nu_0,\nu_1,\nu_2\})\triangleq \mu_{012},
\end{eqnarray}
both of which just involve a single determinant. On the right hand sides,
to make notations simpler, we have introduced a shorthand notation. Likewise,
$\kappa$ for MP4 and MP5 can be written compactly as
\begin{eqnarray}
\kappa^{\rMP 4}&=&\mu_{012}-\mu_{01}\mu_{23}-\mu_{02}\mu_{13}-\mu_{03}\mu_{12},\label{kappa4}\\
\kappa^{\rMP 5}&=&\mu_{01234} \nonumber\\
&-&\mu_{012}\mu_{34} - \mu_{013}\mu_{24} - \mu_{014}\mu_{23} \nonumber\\
&-&\mu_{023}\mu_{14} - \mu_{024}\mu_{13} - \mu_{034}\mu_{12} \nonumber\\
&-&\mu_{01}\mu_{234} - \mu_{02}\mu_{134} -\mu_{03}\mu_{124} - \mu_{04}\mu_{123}.\label{kappa5}
\end{eqnarray}
These expressions are remarkably simpler than the integrands
based on the sum of individual Goldstone diagrams.

\subsubsection{Computational cost for evaluating $\kappa$}
In general, the cost of Eq. \eqref{kappamu} is $O(2^n n^3)$
for computing all the moments $\mu$ involved from principal minors of $\mathbf{G}_{2n+2}^0$,
and $O(nB_n)$ for assembling $\kappa_{n+1}$ from Eq. \eqref{kappamu}, estimated by the number of multiplications. Thus, the use of the determinant trick to sum diagrams, which is a common technique in fermionic QMC for lattice models\cite{rubtsov2004continuous,rubtsov2005continuous}, is essential to avoid the factorial complexity of diagrams at high orders. For $n\ge 6$, Eq. \eqref{kappamu} starts to contain
common intermediates shared by different set partitions, e.g.,
$-\mu_{0123}\mu_{45}+\mu_{01}\mu_{23}\mu_{45}=
-(\mu_{0123}-\mu_{01}\mu_{23})\mu_{45}$.
Then, the recursive algorithm\cite{rossi2017determinant} becomes more advantageous.
Having identified the moment-cumulant relation for $\mu$ and $\kappa$,
the recursive formula can be readily derived by translating the recursive relation\cite{smith1995recursive} between multivariate
moments and cumulants directly,
\begin{eqnarray}
\kappa(\mathcal{I}) = \mu(\mathcal{I})
-\sum_{\mathcal{S}\subset \mathcal{I}',\;\mathcal{S}\ne\emptyset}
\kappa(\{\nu_0\}\cup\mathcal{S})\mu(\mathcal{I}'\backslash\mathcal{S}),\label{kappaN}
\end{eqnarray}
where $\mathcal{I}'\triangleq\mathcal{I}\backslash\{\nu_0\}$, and $\mathcal{S}\ne \emptyset$ comes from the choice of HF reference. This can reduce the cost
for assembling $\kappa_{n+1}$ to $O(3^n)$ asymptotically\cite{rossi2017determinant}.

\section{Monte Carlo algorithm and illustrative example}
After established the formula for MP$n$ correlation energies \eqref{FinalEq},
we now consider its evaluations. The dimensionality of integrations for each $\nu$ in Eq. \eqref{FinalEq} is 9 (including the sum over spins), and the total dimensionality is 9$n$-1 at order $n$, which makes Monte Carlo algorithms a natural choice.
Specifically, at order $n+1$, we define an importance sampling function $p_{n+1}(\nu_0,\cdots,\nu_n)=\prod_{k=0}^{n} p(\nu_k)$ for the configuration $\mathcal{C}_{n+1}=\{\nu_0,\cdots,\nu_n\}$,
\begin{eqnarray}
p(\nu_k)&=&p(\tau_k)p(\sigma_k)p(\sigma_k')p(\vec{r}_k,\vec{r}_k'),\nonumber\\
p(\tau_k)&=&\Delta e^{\Delta\tau},\quad\tau\in(-\infty,0],\nonumber\\
p(\sigma_k)&=&\frac{1}{2},\quad\sigma\in\{\alpha,\beta\},\label{importance}
\end{eqnarray}
where $\Delta=\varepsilon_{\mathrm{LUMO}}-\varepsilon_{\mathrm{HOMO}}$ is the
zeroth-order HF gap of the system. In this work, we investigated two choices for $p(\vec{r}_k,\vec{r}_k')$,
\begin{eqnarray}
p_A(\vec{r}_k,\vec{r}_k')&=&p(\vec{r}_k)p(\vec{r}_k'),\label{schemeA}\\
p_B(\vec{r}_k,\vec{r}_k')&=&\frac{1}{E_{J}|\vec{r}_k-\vec{r}_k'|}p(\vec{r}_k)p(\vec{r}_k'),\label{schemeB}
\end{eqnarray}
where $p(\vec{r})=\frac{1}{K}\sum_{r=1}^{K}|\psi_r(\vec{r})|^2$, $K$ is the dimensionality
of molecular orbitals (MO), and $E_J$ is a normalization factor.
Note that for $\nu_0$, there is no need to generate $\tau_0$.
The two electrons within the same pair
are generated via the standard Markov Chain Monte Carlo (MCMC) method.
Eq. \eqref{FinalEq} is then evaluated simply from the average
$\frac{(-1)^n}{2^{n+1}n!}\left\langle \frac{w_{n+1}\kappa_{n+1}}{p_{n+1}}\right\rangle_{p_{n+1}}$.
The entire algorithm is summarized in Fig. \ref{alg}.

\begin{figure*}
  \centering
  \begin{minipage}{2.0\columnwidth}
  \begin{algorithmic}[1]

  \Function{SMBPT}{$n$}

  \State{Randomly initialize a set of spatial coordinates $\{(\vec{r}_k,\vec{r}_k')\}_{k=0}^{n}$}

  \State{Compute values of molecular orbitals $\{\psi_s(\vec{r})\}_{s=1}^K$ at $\{(\vec{r}_k,\vec{r}_k')\}_{k=0}^{n}$} \Comment{$O(nK^2)$}

  \Loop

    \State{Move $\{(\vec{r}_k,\vec{r}_k')\}_{k=0}^{n}$} randomly to new positions

    \State{Update values of molecular orbitals $\{\psi_s(\vec{r})\}_{s=1}^K$}

    \State{Metropolis update for spatial coordinates according to $p(\vec{r},\vec{r}')$
    in Eq. \eqref{schemeA} or \eqref{schemeB}} \Comment{$O(nK^2)$}

    \If{equilibrated}

        \State{Generate imaginary times and spins
        according to Eq. \eqref{importance} to form a configuration $\mathcal{C}_{n+1}$}

        \State{Construct $\mathbf{G}^0_{2n+2}$ for the given $\mathcal{C}_{n+1}$ from $\{\psi_s(\vec{r})\}_{s=1}^{K}$ and $\{\varepsilon_s\}_{s=1}^{K}$ using Eq. \eqref{GFtImagTime}} \Comment{$O(n^2K)$}

        \State{Evaluate $\kappa_{k+1}$ ($1\le k\le n$) from all principal minors $\mu$ of $\mathbf{G}^0_{2n+2}$ using Eq. \eqref{kappamu} or \eqref{kappaN}} \Comment{$O(f(n))$}

        \State{Compute $\frac{w_{k+1}\kappa_{k+1}}{p_{k+1}}$ and estimates of $E_{k+1}$ for all $1\le k\le n$}

    \EndIf

  \EndLoop

  \EndFunction

  \end{algorithmic}
  \end{minipage}
  \caption{Stochastic MBPT algorithm for correlation energies}\label{alg}
\end{figure*}

The expensive steps in each Monte Carlo step include:
$O(nK)$ for evaluating the values of atomic orbitals (AO) at the sampled spatial points and $O(nK^2)$ for transformation from AO to MO in step 2 (line 7 in Fig. \ref{alg}),
$O(n^2K)$ for constructing $\mathbf{G}^0_{2n+2}$ in step 3 (line 10),
and $O(f(n))$ for evaluating $\kappa_{n+1}$ in step 4 (line 11),
where $f(n)$ is a function depending on a specific numerical scheme
(Eq. \eqref{kappamu} or \eqref{kappaN}) for $\kappa_{n+1}$.
Thus, the total computational cost scales as $O(nK^2+n^2K+f(n))$.
For large systems, assuming $K$ is proportional to the system size $N$,
the present algorithm scales as $O(N^2)$ asymptotically.

To examine the correctness of our formulation in the above sections,
we have implemented the above algorithm in Fig. \ref{alg} in an in-house
program package \texttt{SMBPT}, and studied the prototypical
molecule \ce{H2} with STO-3G at the equilibrium geometry
$R_{\textrm{H-H}}=R_e=0.74144${\AA} and a stretched geometry $R_{\textrm{H-H}}$=4{\AA}
with Eqs. \eqref{schemeA} (scheme A) and \eqref{schemeB} (scheme B).
Albeit being trivial for traditional quantum chemistry methods, this problem is
nontrivial and considerably more complicated than the corresponding two-site Hubbard model
for QMC due to the use of a realistic Coulomb interaction in real space.
The data obtained with sample size being 10$^9$ are shown in Table \ref{dataH2}.
Overall, we found the MP$n$ series can be reproduced by the
present stochastic scheme at both geometries.
Using the summation based on moment-cumulant relations, it successfully extends the previous
MC methods\cite{willow2012stochastic,willow2014stochastic} for MP2 and MP3
to higher orders with a reasonable computational cost.
It can be seen that the scheme B leads to slightly more accurate results than the scheme A.

At higher orders, a new difficulty is found in stochastic evaluations
of the MP$n$ series, which is not obvious in the study of low orders.
From Table \ref{dataH2}, we observed
a rapid growth of variance as $n$ increases, in particular
at the stretched geometry, where the interaction becomes stronger.
This is likely due to both the simplicity of our importance sampling functions
as well as the fermionic sign problem, since $\kappa$ is not always positive.
Therefore, while the obtained data are overall quite encouraging,
further investigations are necessary to fully understand the exact origin of such problem.
Along with other possible improvements, e.g., faster algorithms for computing $\mu$ and assembling $\kappa$, alternative definitions for $\kappa$, as well
as improved sampling techniques, this will be the subject of
a subsequent study.

\begin{table}
\caption{Computed MP$n$ energies $E_n$ for \ce{H2} with STO-3G at the equilibrium geometry
$R_{\textrm{H-H}}=R_e$ and a stretched geometry $R_{\textrm{H-H}}$=4{\AA}
using two different importance sampling functions (scheme A with
\eqref{schemeA} and scheme B with \eqref{schemeB}). The sample size is
$N_{\textrm{MC}}=10^9$.}\label{dataH2}\scriptsize
\begin{tabular}{cccc}
\hline\hline
$n$&	exact  &   scheme A \eqref{schemeA}   &   scheme B \eqref{schemeB} \\
\hline
\multicolumn{4}{c}{$R_{\textrm{H-H}}=R_e$}\\
 1 &	-0.67448	&	-0.67446$\pm$0.00005	&	-0.67450$\pm$0.00004	\\
 2 &	-0.01317	&	-0.01317$\pm$0.00001	&	-0.01317$\pm$0.00001	\\
 3 &	-0.00485	&	-0.00485$\pm$0.00000	&	-0.00485$\pm$0.00000	\\
 4 &	-0.00172	&	-0.00172$\pm$0.00000	&	-0.00172$\pm$0.00000	\\
 5 &	-0.00058	&	-0.00054$\pm$0.00003	&	-0.00059$\pm$0.00001	\\
 6 &	-0.00019	&	-0.00004$\pm$0.00012	&	-0.00022$\pm$0.00008	\\
\multicolumn{4}{c}{$R_{\textrm{H-H}}$=4{\AA}}\\
 1 &	-0.45281	&	-0.45345$\pm$0.00073	&	-0.45282$\pm$0.00001	\\
 2 &	-0.38156	&	-0.38212$\pm$0.00133	&	-0.38210$\pm$0.00108	\\
 3 &	-0.37346	&	-0.37410$\pm$0.00120	&	-0.37353$\pm$0.00125	\\
 4 &	 0.17304	&	 0.17374$\pm$0.00370	& 	 0.17527$\pm$0.00360	\\
 5 & 	 1.22364	&	 1.22205$\pm$0.00882	&	 1.23131$\pm$0.00828	\\
 6 &	 1.22515	&	 1.15467$\pm$0.06289	&	 1.23063$\pm$0.03087	\\
\hline\hline
\end{tabular}
\end{table}

\section{Summary}
In summary, we presented a reformulation of standard MBPT for correlation energies into a general
form \eqref{FinalEq} using Theorems \ref{thm:itf} and \ref{thm:mcr}, which
involves multidimensional integrations that can be evaluated by Monte
Carlo algorithms. The proposed QMC algorithm share similarities with MC-MP2 and MC-MP3\cite{willow2012stochastic,willow2014stochastic}, such as
its low formal scaling $O(N^2)$, which makes it promising for large systems.
The major differences are twofold. First, we use an efficient algorithm based on moment-cumulant relations, which avoids the factorial scaling in using Goldstone diagrams. Second,
in our algorithm all the spatial, spin, and imaginary time variables are sampled stochastically, which are necessary ingredients for high-order perturbation theories.
Like FCIQMC\cite{booth2009fermion} (full configuration interaction quantum Monte Carlo) and AFQMC\cite{motta2018ab} (auxiliary field QMC), the present
QMC algorithm is formulated within an orbital space, but
its evaluation in real space is more similar to standard VMC (variational MC) and DMC.
The advantage of the real-space evaluation is its lower computational scaling and lower requirement for storage with respect to the system size.
However, if full molecular integrals are affordable,
as for small systems, it is also possible to adapt the present QMC algorithm
to the sampling based on molecular integrals as in FCIQMC.
Apart from correlation energies, several other extensions can be readily envisaged, such as the
extension to the finite temperature case and physical properties other than energies.
From a practical aspect, there are still a few obstacles to be overcome
in future. Most importantly, improved sampling methods, along with with ways to alleviate the
fermionic sign problem, need to be developed
in order to apply the stochastic MBPT to large basis sets and systems.
Investigations along these lines are
being carried out in our laboratory.

\section*{Acknowledgements}
Z.L. would like to thank Yunfeng Xiong and Dr. Sihong Shao (Peking University) for helpful discussions
and the Beijing Normal University Startup Package.

\section*{Appendix: Difference between
Eq. (\ref{eFTMBPT}) and similar quantities in finite-temperature MBPT}\label{appendix}
We emphasize that while the expression of $E_{n+1}$ in Eq. \eqref{eFTMBPT} is
in a form similar to that in finite-temperature MBPT (FT-MBPT), this new formula
for correlation energy at a given order does not correspond to any
term in FT-MBPT in the zero-temperature limit. This reflects the
general fact that zero-temperature MBPT (ZT-MBPT) is not the same theory
as FT-MBPT in the zero-temperature limit\cite{kohn1960ground}.
Here, we illustrate the subtle differences between Eq. \eqref{eFTMBPT}
and similar quantities in FT-MBPT in details.
The most distinctive difference is that $\mu$ in Eq. \eqref{eFTMBPT} can take
arbitrary values, while in FT-MBPT it has a physical meaning and controls
the average particle number. However, even if we assume the same $\mu$ is used in
Eq. \eqref{eFTMBPT} and FT-MBPT, Eq. \eqref{eFTMBPT} cannot be derived
from FT-MBPT. For simplicity, we will just discuss the differences for gapped systems.

\begin{figure}\centering
\begin{tabular}{cccc}
\resizebox{!}{0.06\textheight}{\includegraphics{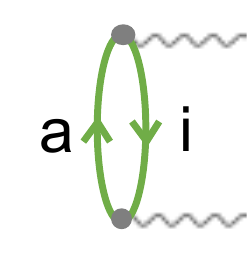}} &
\resizebox{!}{0.06\textheight}{\includegraphics{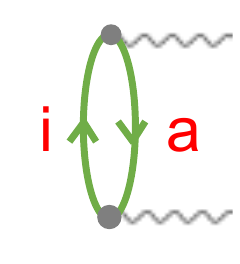}} &
\resizebox{!}{0.1\textheight}{\includegraphics{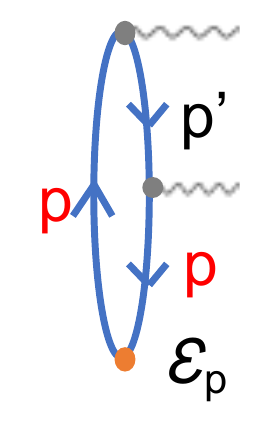}}  &
\resizebox{!}{0.1\textheight}{\includegraphics{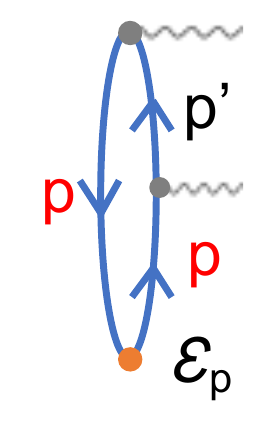}} \\
(a) & (b) & (c) & (d) \\
\end{tabular}
\caption{Diagrams for one-body perturbation at the second order in FT-MBPT.}\label{fig:ftmbpt}
\end{figure}

On the one hand, the corresponding quantity for grand potential
$\Omega_{n+1}$ of a similar form as Eq. \eqref{eFTMBPT} in FT-MBPT reads
\begin{eqnarray}
\Omega_{n+1}=\frac{(-1)^n}{(n+1)!}\int_{0}^{\beta} d\tau_1\cdots d\tau_n
\langle T_\tau[\hV_I(\tau_1)\cdots \hV_I(\tau_n)]\hV\rangle_{c},
\end{eqnarray}
where the factor $1/(n+1)!$ is different from that in Eq. \eqref{eFTMBPT}.
This nontrivial smaller factor precisely cancels the additional contributions from the ensemble average
$\langle\cdots\rangle$,
such that the value of $\Omega_{n+1}$ goes to $E_{n+1}$ in the zero-temperature limit, i.e.,
$\beta\rightarrow\infty$. This can be illustrated by considering a one-body perturbation at the second order,
see Fig. \ref{fig:ftmbpt}. In ZT-MBPT, only Fig. \ref{fig:ftmbpt}(a)
contributes to $E_{2}$, while in FT-MBPT, the additional term Fig. \ref{fig:ftmbpt}(b)
also survives for $\Omega_2$ and has the same value as Fig. \ref{fig:ftmbpt}(a). Only by multiplying
the factor $1/2!$, $\Omega_2$ will become the same as $E_2$ in the zero-temperature limit.

On the other hand, the second order internal energy $U_2$ in FT-MBPT contains two parts,
\begin{eqnarray}
U_{2} &=& \frac{(-1)^1}{1!}\int_{0}^{\beta} d\tau_1
\langle\hV_I(\tau_1)\hV\rangle_{c}\nonumber\\
&+&\frac{(-1)^2}{2!}\int_{0}^{\beta} d\tau_1\int_{0}^{\beta} d\tau_2
\langle T_\tau[\hV_I(\tau_1)\hV_I(\tau_2)]\hH_0\rangle_{c}.\label{eq:U2}
\end{eqnarray}
The first part is similar to $E_2$ in Eq. \eqref{eFTMBPT}, except for the ensemble average.
Thus, it will be twice of $E_2$ as $\beta\rightarrow\infty$, due to the inclusion of both Figs. \ref{fig:ftmbpt}(a)
and (b). Only when the two 'anomalous' diagrams (in the sense that the orbital index $p$ appears both as particles and holes)
from the second part of Eq. \eqref{eq:U2}
are included, see Figs. \ref{fig:ftmbpt}(c) and (d), the additional contribution
in the first part will be cancelled, such that $U_2$ goes to the
same value as $E_2$ in the zero-temperature limit.

In sum, the formula for $E_{n+1}$ \eqref{eFTMBPT} are different from those for $\Omega_{n+1}$
and $U_{n+1}$ in FT-MBPT, even though their values will be the same
in the zero-temperature limit given the same $\mu$.
This novel formula for correlation energies, as a result
of ZT-MBPT followed by a Laplace transformation to introduce an artificial imaginary time,
cannot be obtained from any physical quantity in FT-MBPT by taking in the zero-temperature limit.


%

\end{document}